\documentclass[11pt,twoside]{article}
\usepackage{amsmath,amssymb,graphicx,bm,epsfig,color}
\usepackage{geometry}             
\usepackage{qcircuit}
\usepackage{physics}
\usepackage{subfig}

\begin{document}
\title{{\bf  NISQ computing for decision making under uncertainty}}

\author{Names}
\author{H.W.L. Naus\thanks{E-mail: rik.naus@tno.nl}\\
Quantum Technology, TNO\\
PO Box 155, 2600 AD Delft, Netherlands}

\date{\today}
\maketitle

\begin{abstract}
Noisy Intermediate-Scale Quantum computers are expected to be available this year.
It is proposed to exploit such a device for decision making under uncertainty. 
The probabilistic character of quantum mechanics reflects this uncertainty.
Concomitantly, the noise may add to it. The approach is standard in the sense that Bayes decision
rule is used to decide on the basis of maximum expected reward. The novelty is to model
the various action profiles and the development of `nature' as unitary transformations on a 
set of qubits. Measurement eventually yields samples of classical binary random variables
in which the reward function has to be expressed. In order to achieve sufficiently low
variances for reliable decision making more runs of such a quantum algorithm are necessary.
Some simple examples have been worked out to elucidate the idea. Here the  calculations
are still analytically feasible. Presently lacking an operating
quantum device, the QX simulator of Quantum Inspire has been used to
generate the necessary samples for comparison and demonstration. First obtained results
are promising and point at a possible useful application for noisy intermediate-scale
 quantum computers.
\end{abstract}

\newcommand{\tf}{\tfrac{1}{2}\sqrt{2}}


\section{Introduction}

Large-scale universal quantum computers \cite{NC} allowing fault tolerant quantum computing \cite{Fowler}
are not foreseen in the near future. However, Noisy
Intermediate-Scale Quantum (NISQ) devices will be available in the next years \cite{Preskill}. 
Such quantum computers have a limited number of qubits, say up to 100 and suffer from
noise which reduces the fidelity of the quantum gates. A surface code to correct errors 
\cite{Fowler} cannot be implemented. 
The question therefore arises how to usefully apply NISQ technology. In the seminal
paper \cite{Preskill} such opportunities, especially for computing, have been discussed.

This note proposes another application for NISQ computers, that is decision making
in the face of uncertainty \cite{OR}. Classical probabilities and, more generally,
probability theory are of course the essential guidelines in the process of decision
making if uncertainties are involved. Probability density functions and their possible updates
have been used in practice, see, e.g., \cite{Gio}.
In our proposal we do not choose probability distributions,
but exploit quantum mechanical probabilities. In essence, we model the problem in
terms of the state\footnote{This is, of course, not a quantum state} of nature,
typically with a uncertain evolution,
and in terms of different actions (alternatives). The latter, in combination with the occurring state
of nature, may give different rewards (payoffs). These actions and  the evolution of the
state of nature are modeled quantum mechanically by
means a set of unitary transformations on a limited number of qubits. The latter are supposed
to be initialized in their ground states as is commonly assumed in quantum algorithms. 
Eventually, we measure the product of all $\sigma_z$ operators of the qubits;
usually but somewhat sloppy this is called `measuring the qubits'. 
The measurement yields a number of classical bits, corresponding to one sample.
Repeating the calculation, including the measurements, $N$ times then yields $N$ samples
which are used to estimate the expectation value of the reward (payoff).      
The Bayesian decision rule \cite{OR} maximizes this expected reward.

Of course, we need to motivate our choice for a NISQ device in decision making.
First, we expect that even for complex decision making problems the necessary number
of qubits is well below the expected limit of 100 qubits. Although the algorithms are
deterministic, the outcome of the measurements is inherently probabilistic. In this
way, the uncertainties are simulated in a natural way. The noise in a NISQ computer
obviously also reflects these uncertainties and is therefore no nuisance like in other applications.
Updating probabilities and probability density functions on the basis
of previous experimentation \cite{OR} and/or new acquired information \cite{Gio}
is conveniently implemented by means of controlled quantum operations like a CNOT gate.
Note that all two-qubit gates may introduce entanglement in the system, in this way
modeling correlated random variables.   
We emphasize, however, that the merits of this method need eventually to be demonstrated
in practice. Two advantages can be envisaged. Since the quantum technique is principally
different from the classical one, it may be more reliable and robust. The involved
quantum sampling may also speed up the process of classical decision making where
time consuming Monte Carlo simulations are often necessary. 

At present, a NISQ computer is not yet available. Quantum simulators, however, are
already available. In our examples, we have used the QX simulator of Quantum Inspire
\cite{QI}. As of yet, the simulator supports only one error model which may be used
to include the imperfections of a real device.  Quantum Inspire actually aims for
a spin-qubit quantum computer in the very nearby future.

There is an obvious relation with quantum sampling where a quantum computer is exploited
to generate samples of a probability density function. Here we also generate samples to
obtain classical bits in which the reward function is expressed. 
No guiding classical probability distribution is however used; the guiding principle is
the combination of the uncertain state of nature and the various actions.
This needs to be modeled in by unitary transformations on the multi-qubit state.
The method has also some resemblance with quantum random walks \cite{Kempe}.     

The outline of this note is as follows. First, we very briefly introduce the
necessary concepts of decision making. Next we fix our notation for quantum computing.
In section \ref{Concept} the general concept of exploiting quantum computing
in decision making is presented. It is followed by sections working out
three simple examples with increasing complexity. Finally,
some conclusions and an outlook are presented.  

\section{Preliminaries}
In this section we start by shortly introducing the few aspects of decision analysis
and theory which we actually want to use. The more interested reader may consult
the standard textbook \cite{OR} and/or the recent review \cite{Chal}.
Our work is also based on an approach with classical probability density functions and Bayesian updates \cite{Naus}.
 
\subsection{Decision making}
The simplest example of decision making under uncertainty may be to decide whether or not
to take an umbrella while going out. Such a decision is typically based on a weather forecast.
It contains the essential ingredients which are several actions or alternatives to choose from,
an eventual `state of nature'
and a resulting reward or payoff. In the example the possible eventual states of nature 
are rain or no rain. The reward is the difference between value, dry versus soaked clothes,
and cost, carrying the umbrella. These concepts need to be quantified and often
monetary values are used\footnote{The concept utility goes beyond this.}.
Possibly unsubstantiated and subjective probabilities or probability distributions are invoked
for predicting the state of nature.
The decision maker chooses the optimal action based on some criterion.
In \cite{OR}, three such criteria are given:
\begin{itemize}
	\item Maximin payoff\\
	  For each action find the minimum reward over all states. Next, find the maximum of these
minimum payoffs.  Choose the action whose minimum payoff yields this maximum. Obviously, it is 
	  a pessimistic view and is very cautious.
	  \item Maximum likelihood \\
	  Identify the most likely state of nature, that is with the highest prior probability.
For this state choose the decision with the highest reward. 
	  This criterion excludes low-probability high-payoff gambling.
	  \item Bayes decision rule \\
	  Use the best available estimates of the probabilities (or probability distributions)
	  of the respective states and calculate the expected reward for the various actions.
Choose the one with the maximum reward. 
\end{itemize}
In this study we restrict ourselves to the latter criterion, that is Bayes decision rule.
It obviously requires more computations (simulations) then the other two.

Up to now, we have described decision making without experimentation.
It can be extended to include experimentation \cite{OR} or Bayesian(-like) updates based on the already
acquired information. The aim is to improve the estimated prior probabilities (or density functions).
The improvements are called posterior probabilities (or updated density functions \cite{Naus}).
For simple examples we refer to \cite{OR,Naus}.

\subsection{Notation quantum mechanics}
In this section we define the notation to describe the necessary NISQ quantum mechanical
part. For a single qubit we use as standard or computational basis
\begin{equation}
\ket{0} =
\begin{pmatrix}
1 \\
0
\end{pmatrix},
  \ket{1} = 
\begin{pmatrix}
0 \\
1
\end{pmatrix}.
\end{equation}
It is straightforwardly extended to more qubits; for two qubits we explicity have
\begin{equation}
\ket{00} =
\begin{pmatrix}
1 \\
0 \\
0 \\
0
\end{pmatrix},
  \ket{01} = 
\begin{pmatrix}
0 \\
1 \\
0 \\
0
\end{pmatrix},
  \ket{10} = 
\begin{pmatrix}
0 \\
0 \\
1 \\
0
\end{pmatrix},
  \ket{11} = 
\begin{pmatrix}
0 \\
0 \\
0 \\
1
\end{pmatrix}.
\end{equation}
The initial state for $n$ qubits will always be chosen as $\ket{0}^{\otimes n}$.
Note that the computational states are the eigenstates of the Pauli $\sigma_z$ (or $Z$) operators
\begin{equation}
\sigma_z\ket{0} = \ket{0}, \quad
\sigma_z \ket{1} = -\ket{1}.
\end{equation}
The eigenvalues $1,-1$ are the possible results of a measurement of $Z$. They correspond
to classical bits $0,1$, thereby refering to the eigenstate. The following
NISQ algorithms are always terminated by a simultaneous measurement of
$Z^{[1]}, Z^{[2]} \cdots$ and $Z^{[n]}$. In this way a classical bit vector
$\vec{s}$ with $s_k=0,1; k=1, \cdots n$ is generated. 
Because of the inherent probabilistic
character of quantum physics, it is produced with some probability
depending on the  state just before the measurements. This deterministic state
follows from applying unitary operations on the initial state. Repeating the whole
procedure, that is including the final measurement, then yields samples of the classical,
random binary vector $\vec s$.  

\section{Concept}\label{Concept}
The leads us to the concept of decision making under uncertainty by means of a NISQ device.
For each possible action $A_j$, the reward function has to expressed in
the classical random binary variables $\vec s$:
\begin{equation}
r_j(\vec s)=v_j(\vec s)-c_j(\vec s)
\end{equation}
with value function $v(\vec s)$ and cost function $c(\vec s)$. These functions
may be different for the various actions, but are not necessarily so. Furthermore,
they may depend on other deterministic variables or parameters. The uncertainty 
is simulated by obtaining the random samples $\vec s$ by means of the NISQ computer.
Apart from the quantum probabilities, the latter actually adds some uncertainty because
of noise. 

For each action, a quantum algorithm in terms of unitary operations $U_k$ has to
implemented. Usually also the evolution of the state of nature is mimicked by
an unitary  transformation $U(\tau)$, where $\tau$ is a evolution parameter like a 
dimensionless time. For example, we get the final state
\begin{equation}
| \psi \rangle  = U \, \mathcal U(\tau) \ket{0}^{\otimes n}.
\label{eq:U}
\end{equation} 
 The resulting algorithms have to be performed a number, say $N$,
of times. In this way, $N$ samples of the classical bits are obtained for each action.
Inserting the samples into the reward function eventually yields samples of various
rewards and therefore also their averages as sample means.
These are used as decision criterion,
the action with the highest expected reward is selected. Concomitantly, it is
possible to estimate the variances of the rewards function by their sample variance.
It of course yields an impression how sensitive a decision may be for the uncertainties.
Specifally, we obtain for each action the variates $\vec s_k, k=1, \cdots, N$ and thus 
the samples $r(\vec s_k)$. The estimated expected reward\footnote{Below we include a subindex to indicate the action, {\it i.e.}, $E_j$ denotes the expected reward for action $A_j$.}
$E$ follows as
\begin{equation}
E[r]  = \frac{1}{N} \sum_{k=1}^N r(\vec s_k).
\end{equation} 
The output is sometimes provided as a list of found different samples $\vec s(i), i=1, L$
and their fractions $n_i/N$ which is an estimate of their probabilities $p_i$.
In that case, we obtain as average
\begin{equation}
E[r] =\sum_{i=1}^L \frac{n_i}{N}  r(\vec s(i)) =\sum_{i=1}^L p_i r(\vec s(i)).
\end{equation} 
This is especially convenient if the number of qubits is small and, consequently,
the number of different variates is not that large. 
Recall that we do not yet have a NISQ device at our disposal and therefore use the 
QX simulator of Quantum Inspire.

Although this concept appears to be simple, some thinking is required if a practical
decision making needs to be solved. First, the set of possible actions has to be identified.
Next, one needs to define appropriate reward
functions in terms of the binary random variables $\vec s$ and other parameters. Obviously,
this resembles the approach in classical decision making. Note that there been published
a vast amount of literature on this subject. Exploiting a quantum computer requires
that the evolution of the state of nature and the consequences of the various actions
need to be expressed as unitary transformations. For the case of decision making `with 
experimentation' we propose to apply controlled unitary two-qubit gates, for example
the CNOT.  
This leads to a generalized form of eq.(\ref{eq:U}) 
\begin{equation}
| \psi \rangle  = U_m \, \mathcal U(\tau_m,\tau_{m-1})\,
 U_{m-1} \, \mathcal U(\tau_{m-1},\tau_{m-2})\cdots
 U_{1} \, \mathcal U(\tau_{1},\tau_{0}) \ket{0}^{\otimes n},
\end{equation} 
where `time' is ordered: $\tau_m > \tau_{m-1} > \cdots > \tau_0$. 
The unitary operations $U_j, j=1, \cdots, m$ can be controlled gates.

\section{Taking an umbrella} 
The decision to be made is (not) taking an umbrella, given a certain
probability $p_r$ of rain. The umbrella, however, is heavy -- so carrying it has a price. 
The possible states of nature can be coded in two classical bits: $s_0=0, 1$ corresponding
to rain, no rain and $s_1=0, 1$ corresponding to travelling light, heavy.
The quantum implementation therefore needs two qubits, {\it i.e.},
\begin{align}
\text{qubit} [0] : \quad &\ket{0} \quad \text{no rain}, &\ket{1}& \quad \text{rain}, \nonumber\\
\text{qubit} [1] : \quad &\ket{0} \quad \text{light}, &\ket{1}& \quad \text{heavy}.
\end{align}
The reward functions therefore depend on the two binary variables $s_0, s_1$.  
\subsection{Simplest case}
\begin{figure}[h!]
\centering
	\subfloat{
	\scalebox{1.75}{
		\begin{tabular}{c}
\Qcircuit @C=1em @R=.8em
{\lstick{\ket{0}}  & \gate{R_y(\tau)} & \meter \\
\lstick{\ket{0}}  & \qw & \meter} 
		\end{tabular}
}
}
	\qquad \qquad
	\subfloat{
		\scalebox{1.75}{
		\begin{tabular}{c}
\Qcircuit @C=1em @R=.8em
{\lstick{\ket{0}}  & \gate{R_y(\tau)} & \meter \\
\lstick{\ket{0}}   & \gate{X} &  \meter} 
		\end{tabular}
}
}
\caption{Corresponding quantum circuits, 
l.h.s: action 1; r.h.s: action 2} \label{cir:1} 
\end{figure}
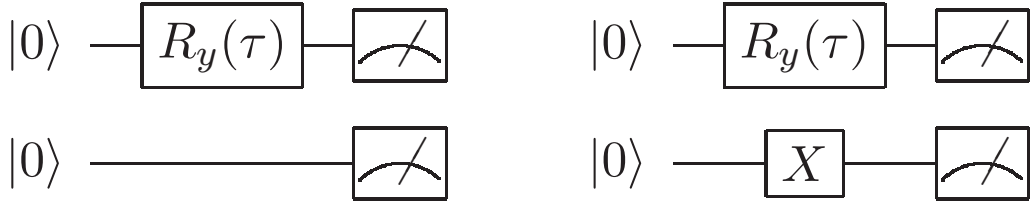

In the simplest case, we only consider two actions, action 1 is {\em not} taking the umbrella
whereas action 2 is taking the umbrella. Let us define the value function by noting that is
some value $v$ to remain dry which does not happen if $s_0=1$ and $s_1=0$;
so we take
\begin{equation}
v(s_0,s_1)=v(1-s_0(1-s_1)).
\end{equation}
Because the umbrella is heavy the cost is chosen as
\begin{equation}
c(s_0,s_1)=c s_1,
\end{equation}
which is independent of $s_0$. 
These functions apply for both decisions, so we also have only one reward function
 in this case 
\begin{equation}
r(s_0,s_1)=v(1-s_0(1-s_1))-c s_1.
\end{equation}
The weather development is described with the unitary one-qubit gate 
\begin{equation}
R_y(\tau) \ket{0} = \cos{(\frac{\tau}{2})}\ket{0} -\sin{(\frac{\tau}{2})}\ket{1},
\end{equation}
that is a $y$-rotation around $\tau$. It yields a probability of rain 
\begin{equation}
p_r = \sin^2{(\frac{\tau}{2})},
\label{eq:pr}
\end{equation}
so $\tau$ can be chosen to match the prediction. 
Despite the subjective experience, rain versus no rain does not depend on the
decision\footnote{Excluding `butterfly effects'.}.
The complete unitary operation for action 1 is given by the rotation on the $q[0]$
and the identity on $q[1]$, resulting in the state
\begin{equation}
|\Psi_1\rangle = (R_y^{[0]}(\tau) \otimes \mathcal I^{[1]}) \ket{00}.  
\end{equation}
Taking the umbrella corresponds to an $X$-operation on $q[1]$, yielding 
\begin{equation}
|\Psi_2\rangle = (R_y^{[0]}(\tau) \otimes \sigma_x^{[1]}) \ket{00}.  
\end{equation}
The corresponding complete circuits, {\it i.e.}, including the measurements are shown in Figure (\ref{cir:1}).

This example is that simple that we can proceed analytically. Standard quantum mechanical
calculations yields for action 1
\begin{align}
p(s_0=0, s_1=0) &=   \cos^2{(\frac{\tau}{2})}, \quad p(s_0=1,s_1=0) = \sin^2{(\frac{\tau}{2})}, \nonumber\\
p(s_0=0, s_1=1), &=   p(s_0=1,s_1=1) = 0,
\end{align}
and eventually the expected reward 
\begin{equation}
E_1[r]= v \cos^2{(\frac{\tau}{2})}.
\end{equation}
The analogous computations for action 2 give
\begin{align}
p(s_0=0, s_1=1) &=   \cos^2{(\frac{\tau}{2})}, \quad p(s_1=1,s_1=1) = \sin^2{(\frac{\tau}{2})}, \nonumber\\
p(s_0=0, s_1=0), &=   p(s_0=1,s_1=0) = 0,
\end{align}
with  expected reward 
\begin{equation}
E_2[r]= v-c.
\end{equation}
Given the probability of rain $p_r$ (\ref{eq:pr}),
one can now take the `best' decision. 
It is taking the umbrella for $p_r > c/v$.

We nevertheless have implemented this example in the QX simulator to generate
samples of $\vec s$.  For each action 1024 runs have been done, {\em i.e.}, $N=1024$.
The resulting expected rewards are consistent with the analytical results.

\subsection{With experimentation - entanglement}
\begin{figure}[h!]
\centering
	\scalebox{2}{
		\begin{tabular}{c}
\Qcircuit @C=1.em @R=.8em
{\lstick{\ket{0}} & \gate{R_y(\tau_0)} & \ctrl{1} & \gate{R_y(\tau-\tau_0)} & \qw & \meter \\
\lstick{\ket{0}} & \qw & \targ & \qw &  \qw & \meter} 
		\end{tabular}
}
\caption{Circuits} \label{cir:2} 
\end{figure}
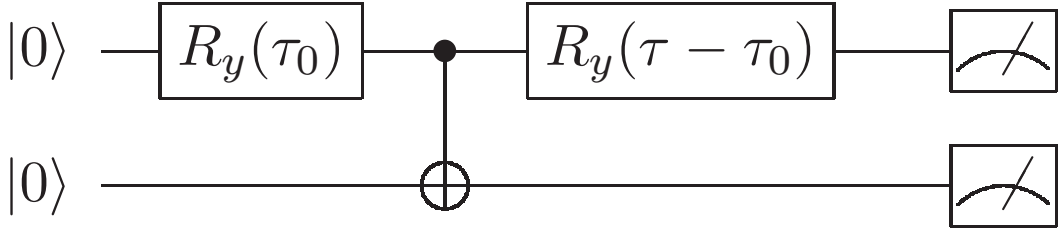

We extend this example by adding an third action. It consists out of waiting a certain time
and then taking the decision also on the basis of the weather development thus far.
An obvious criterion is rain at that moment\footnote{Alternatives like an updated weather
forecast are also reasonable.}. 
A quantum mechanical implementation of this action is
\begin{equation}
|\Psi_3\rangle = (R_y^{[0]}(\tau-\tau_0) \otimes \mathcal I^{[1]}) \, C_N \,
(R_y^{[0]}(\tau_0) \otimes \mathcal I^{[1]}) \ket{00},  
\end{equation}
where $C_N$ denotes the entangling CNOT gate. It corresponds to the quantum circuit shown in Figure (\ref{cir:2}).
It is furthermore assumed that waiting has its price and for this decision the cost 
function is therefore taken as
\begin{equation}
c_3(s_0,s_1)= c s_1 +d.
\end{equation}
Since the cost increases for longer waiting we take $d \propto \tau_0/\tau$.
Although the calculations are somewhat more involved, one can still proceed analytically.
The resulting probabilities follow as
\begin{align}
p(s_0=0, s_1=0) &=   \cos^2{(\frac{\tau-\tau_0}{2})} \cos^2{(\frac{\tau_0}{2})}, \nonumber \\
p(s_0=0, s_1=1) &=   \sin^2{(\frac{\tau-\tau_0}{2})} \sin^2{(\frac{\tau_0}{2})} , \nonumber\\
p(s_0=1, s_1=0) &=   \sin^2{(\frac{\tau-\tau_0}{2})} \cos^2{(\frac{\tau_0}{2})} , \nonumber\\
p(s_0=1, s_1=1) &=   \cos^2{(\frac{\tau-\tau_0}{2})} \sin^2{(\frac{\tau_0}{2})} , 
\end{align}
eventually yielding the  expected reward 
\begin{eqnarray}
E_3[r] &=& v\left\{(\cos^2{(\frac{\tau-\tau_0}{2})} \cos^2{(\frac{\tau_0}{2})} + \sin^2{(\frac{\tau_0}{2})}\right\}
- c \sin^2{(\frac{\tau_0}{2})}  - d \nonumber \\
&=& v\left\{1 - \sin^2{(\frac{\tau-\tau_0}{2})} \cos^2{(\frac{\tau_0}{2})}\right\}
- c \sin^2{(\frac{\tau_0}{2})}  - d.
\label{eq:ER1}
\end{eqnarray}
Once again, we also have performed QX simulations and have approximately reproduced the
analytical probabilities for the third action as well.
 
\begin{figure}[htb]
\centering
\includegraphics[width=7.75cm]{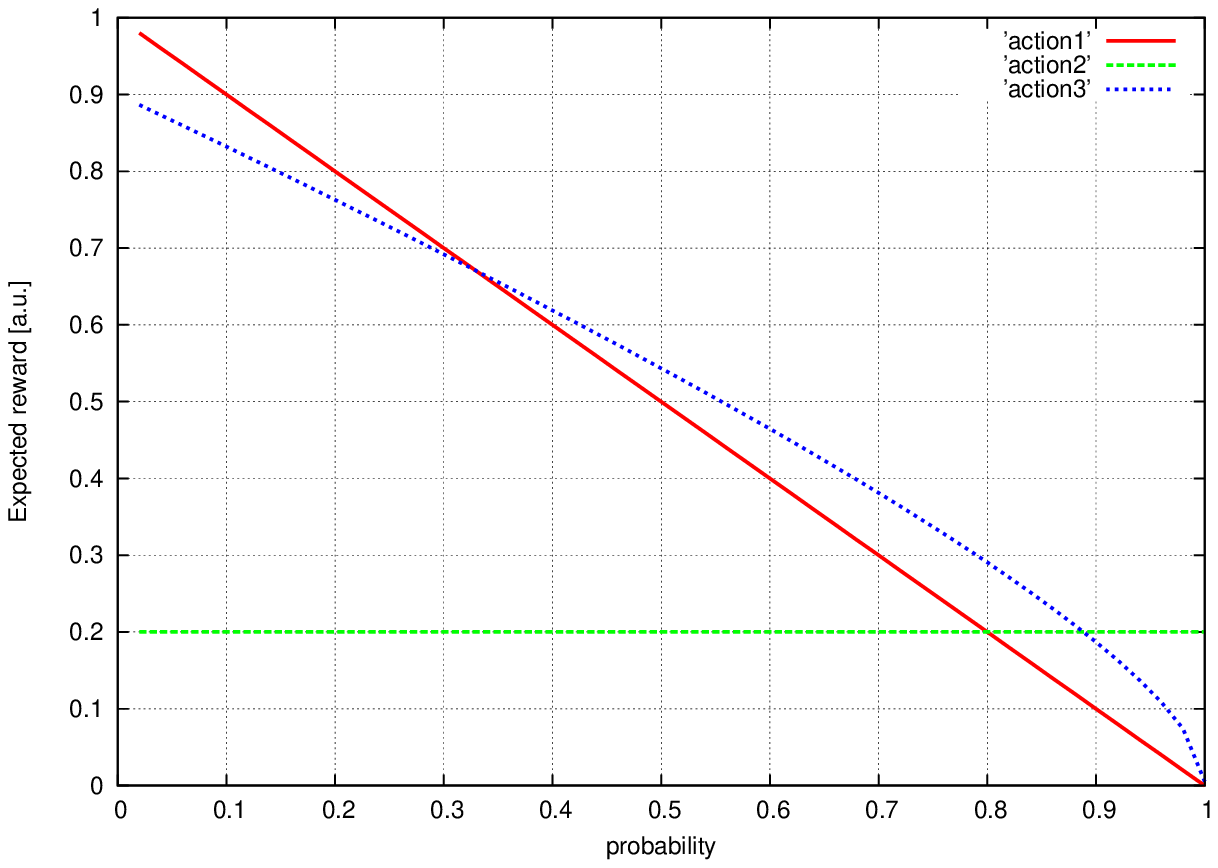}
\includegraphics[width=7.75cm]{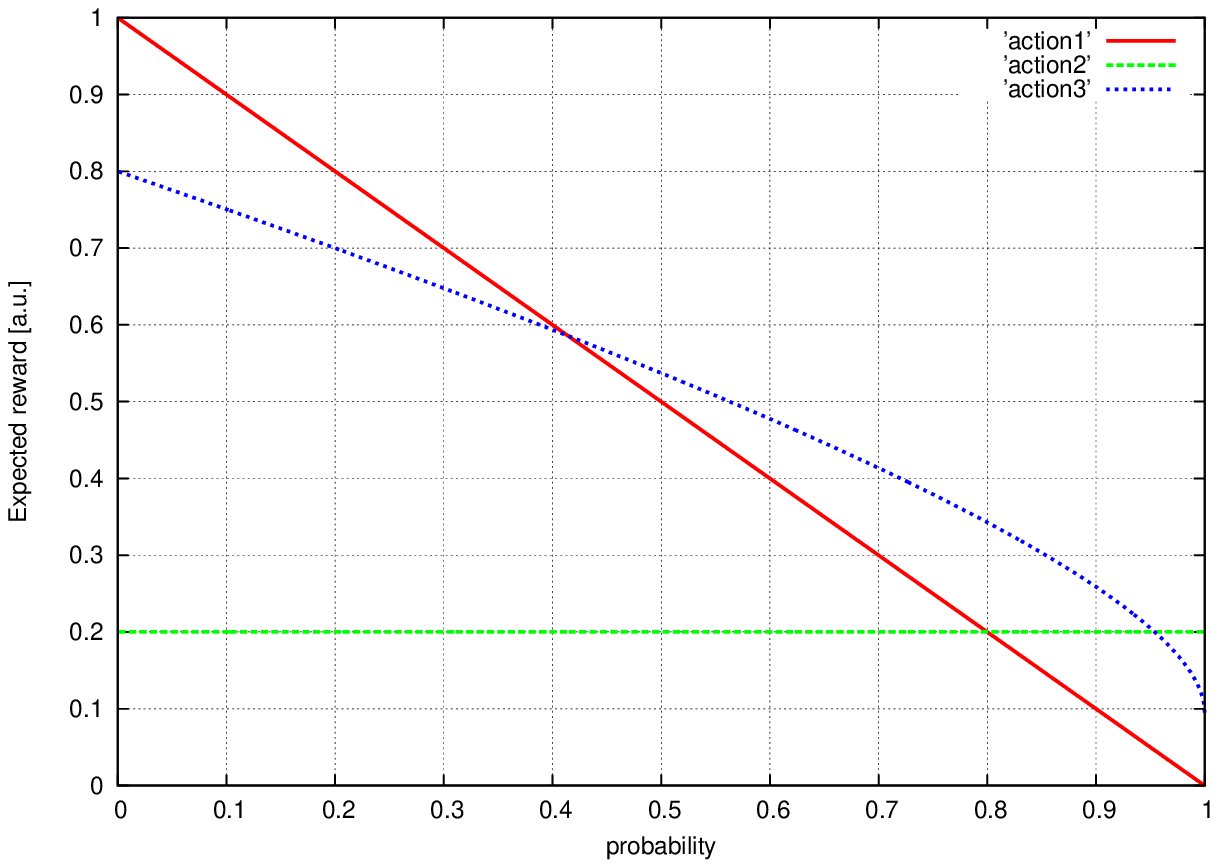}
\caption{Expected rewards for the three different actions as a function of the rain probability $p_r$.
L.h.s. $\tau_0=0.2\tau, d=0.1$; r.h.s $\tau_0=0.4\tau, d=0.2$.  \label{fig:1}}
\end{figure}

If all parameters in the reward functions and in the unitary transformations are fixed,
the decision is taken by selecting the one with maximal reward. It may illustrative to show
the expected reward as a function of one the parameters. A natural choice is the probability of rain
$p_r$, which determines the `time' (rotation angle) $\tau$. We fix the additional cost as
$d= \frac{\tau_0}{2\tau}$ and take as parameters $v=1.0, c=0.8$.
Figure (\ref{fig:1}) depicts the rewards for the three possible actions and two waiting times.

\subsection{Including variances}
It is also possible to analytically derive the variances\footnote{As is well known
the standard deviation $\sigma$ is the square root of the variance.}
of the expected rewards for the various actions. 
In case of only sampling, either by means of the QX simulator or a NISQ computer, these
quantities are replaced by sample variances.
For action 1 we get
\begin{equation}
E_1[r^2]= v^2 \cos^2{(\frac{\tau}{2})} \; \Rightarrow \; \sigma_1=\tfrac{1}{2}v\sin{\tau}.
\end{equation}
The variance of the expected reward for action vanishes identically. Sampling
with a NISQ device will yield a finite but hopefully small variance.
The expected squared reward for action 3 is obtained as
\begin{align}
E_3[r^2] &= (v-d)^2 \cos^2{(\frac{\tau-\tau_0}{2})} \cos^2{(\frac{\tau_0}{2})} \nonumber\\
 &+ (v-c-d)^2 \sin^2{(\frac{\tau_0}{2})}
 + d^2 \sin^2{(\frac{\tau-\tau_0}{2})} \cos^2{(\frac{\tau_0}{2})}.
\end{align}
The variance follows from $\sigma_3^2=E_3[r^2]-(E_3[r])^2$, cf. (\ref{eq:ER1}), but
we omit this rather lengthy expression.

\begin{figure}[htb]
\centering
\includegraphics[width=7.75cm]{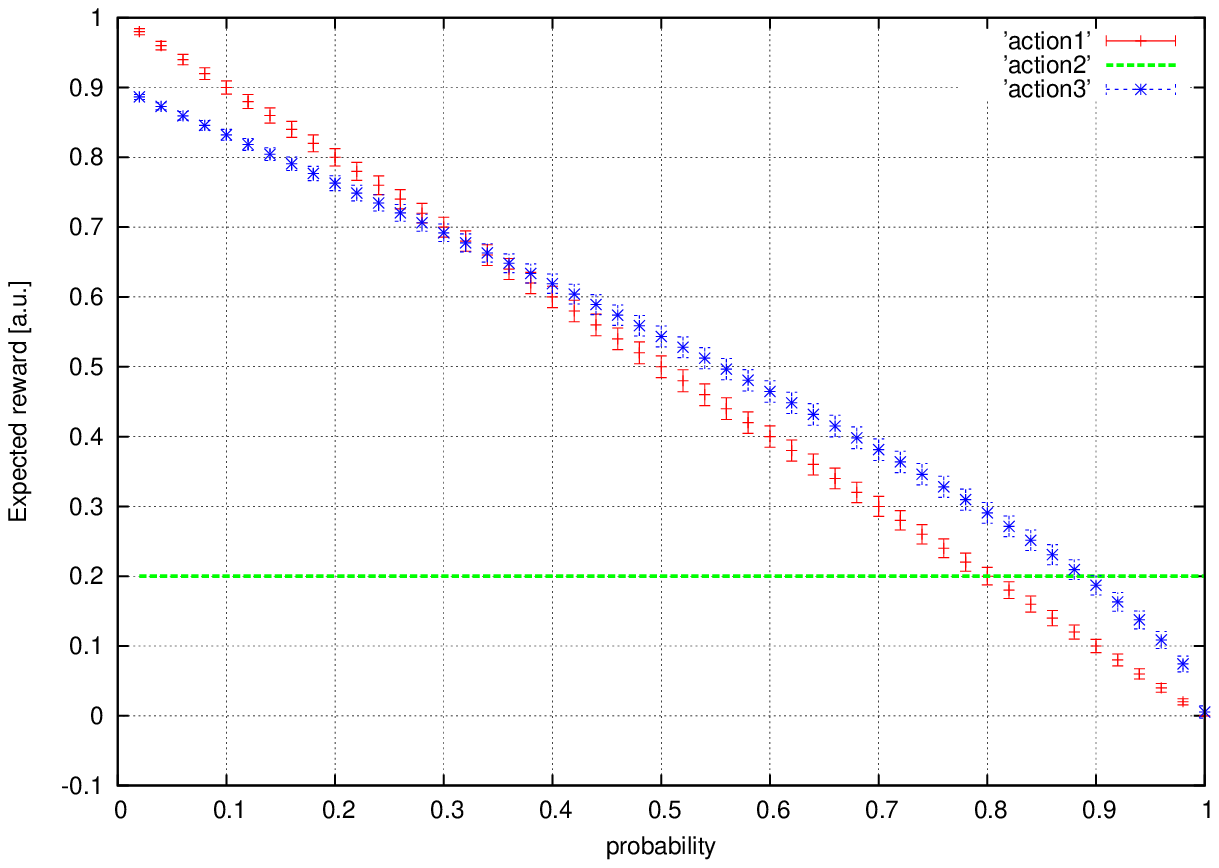}
\includegraphics[width=7.75cm]{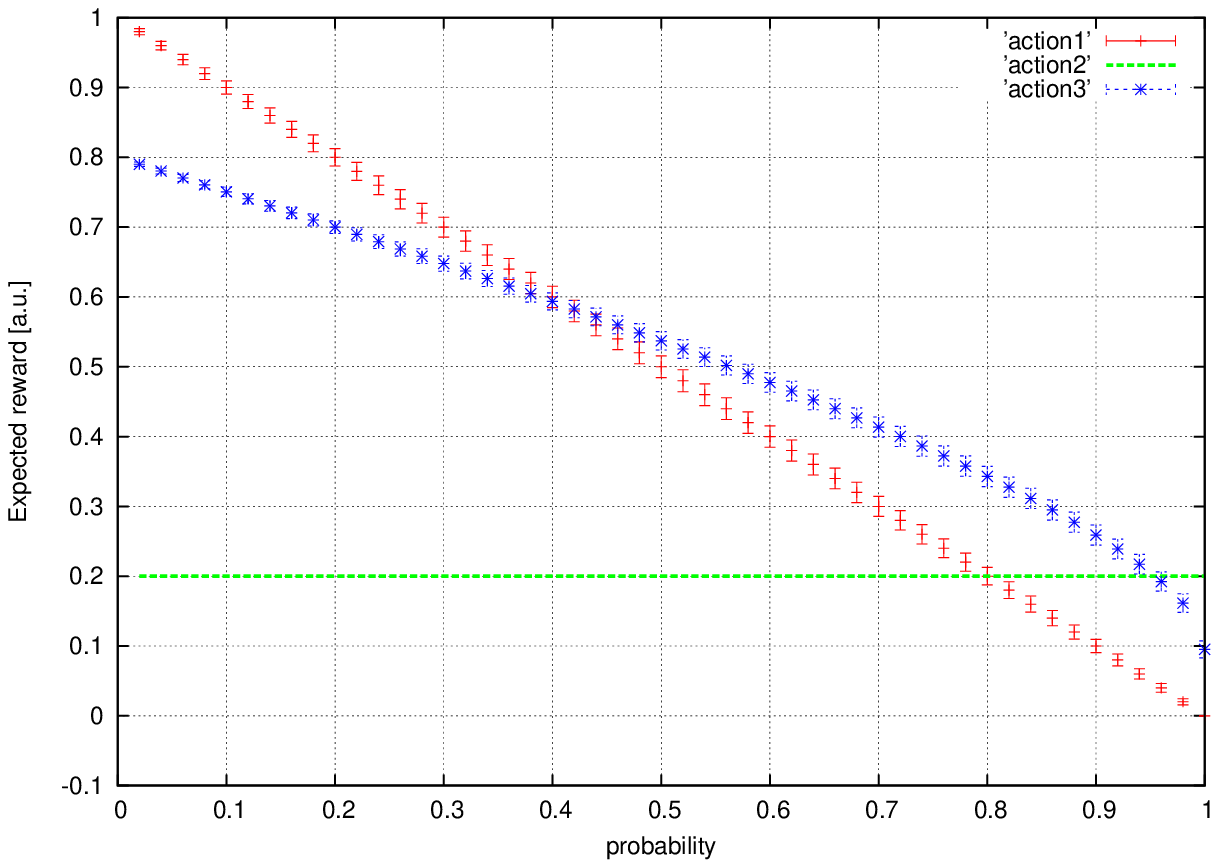}
\caption{Expected rewards and variances for the three different actions as a function of
the rain probability $p_r$; $N=1024$.
L.h.s. $\tau_0=0.2\tau, d=0.1$; r.h.s $\tau_0=0.4\tau, d=0.2$.  \label{fig:2}}
\end{figure}

It is clear, however, that the resulting standard deviation is of the same order of magnitude as
the expected reward. Consequently, it is impossible to make a decision based on one experiment/run.
Performing $N$ runs with the QX simulator, and eventually on the NISQ device as well, reduces
the standard deviation by a factor of $1/\sqrt{N}$. Figure (\ref{fig:2}) shows the 
expected reward including standard deviation calculated for $N=1024$, at present
the maximum number of runs for the QX simulator.
We see that sensible decision making is still possible. The variance for the results
generated with a NISQ computer may be larger.

\section{Example: entangling nature}
\subsection{Scenario and model}
In this section we extend and modify the previous example. Apart from considering
the possibility of rain the temperature is considered as well. If the temperature is high then one
can leave a jacket at home, which is convenient.
However, in case of low temperatures a jacket
is necessary. Unfortunately, it is inconvenient to carry.
Hence one needs to decide about taking an umbrella {\em and} carrying the jacket.
The possible states of nature are now coded in four classical bits: $s_0=0, 1$ corresponding
to rain and  no rain,
$s_1=0, 1$ corresponding to warm and cold,
$s_2=0, 1$ corresponding to travelling light and heavy and
$s_3=0, 1$ corresponding to travelling conveniently and inconveniently.
The quantum implementation therefore needs four qubits, {\it i.e.},
\begin{align}
\text{qubit} [0] : \quad &\ket{0} \quad \text{no rain}, &\ket{1}& \quad \text{rain}, \nonumber\\
\text{qubit} [1] : \quad &\ket{0} \quad \text{warm}, &\ket{1}& \quad \text{cold}, \nonumber\\
\text{qubit} [2] : \quad &\ket{0} \quad \text{light}, &\ket{1}& \quad \text{heavy}, \nonumber \\
\text{qubit} [3] : \quad &\ket{0} \quad \text{convenient}, &\ket{1}& \quad \text{inconvenient}. 
\end{align}
The reward functions therefore depend on the four binary variables $s_0, s_1, s_2, s_3$.  
Analogously to the previous example, we define the value function
\begin{equation}
v(s_0,s_1,s_2,s_3)=v_0(1-s_0(1-s_2)) +v_1(1-s_1(1-s_3)).
\end{equation}
Since inconvenience has its price, the cost is extended as
\begin{equation}
c(s_0,s_1,s_2,s_3)=c_0 s_2 + c_1 s_3.
\end{equation}
The reward remains value minus cost.
Just as above the cost function has to be adapted for one action;
we will get back to this point.

First, we further modify the scenario by assuming that the decision cannot be
based on the weather forecast but only on experience. That tells us that the conditions
are either ``warm and dry" {\it or} ``cold and rainy", with about 50\% probability.
Such a (weather) state of nature corresponds with the following Bell state\footnote{Of course,
the Bell state with the relative $-$sign also does the job.} of qubits $[0]$ and $[1]$
\begin{equation}
|\Phi^+\rangle = \tfrac{1}{2}\sqrt{2}(\ket{00} + \ket{11}).
\end{equation}
The unitary operation $C_N H^{[0]}$ transforms $\ket{00}$ to $|\Phi^+\rangle$ and,
consequently, we define the fixed\footnote{No `time' parameter is required here.}
 `weather evolution operator' as 
\begin{equation}
U_W = (\mathcal I^{[3]} \otimes \mathcal I^{[2]} \otimes C_N^{[01]})      
\, (\mathcal I^{[3]} \otimes \mathcal I^{[2]} \otimes \mathcal I^{[1]} \otimes  H^{[0]})
= U_W^{[01]} = C_N^{[01]} \, H^{[0]},
\end{equation}
where $H$ is the Hadamard gate. Note the introduction of a shorter notation
by only indicating the non-trivial operations and thus omitting identity gates. 

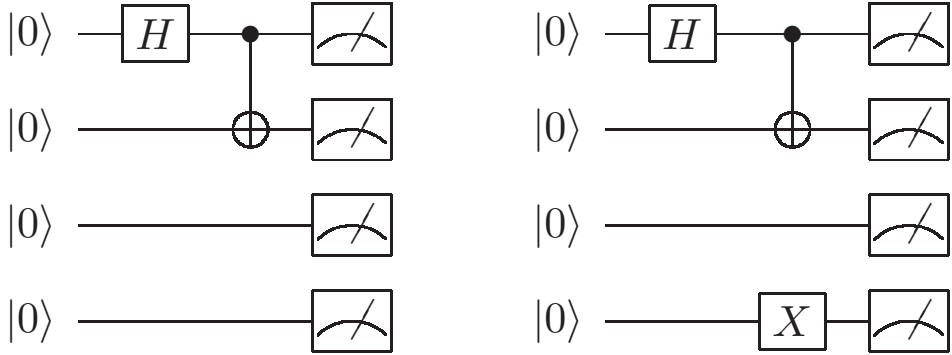
\begin{figure}[h!]
\centering
	\subfloat{
	\scalebox{1.5}{
		\begin{tabular}{c}
\Qcircuit @C=1em @R=.8em
{\lstick{\ket{0}}  & \gate{H} & \ctrl{1} & \meter \\
\lstick{\ket{0}}  & \qw & \targ &  \meter  \\
\lstick{\ket{0}}  & \qw & \qw & \meter  \\
\lstick{\ket{0}}  & \qw & \qw & \meter} 
		\end{tabular}
}
}
	\qquad \qquad
	\subfloat{
		\scalebox{1.5}{
		\begin{tabular}{c}
\Qcircuit @C=1em @R=.8em
{\lstick{\ket{0}}  & \gate{H} & \ctrl{1} & \meter \\
\lstick{\ket{0}}  & \qw & \targ & \meter  \\
\lstick{\ket{0}}  & \qw & \qw & \meter  \\
\lstick{\ket{0}} & \qw  & \gate{X} &  \meter} 
		\end{tabular}
}
}
\caption{Corresponding quantum circuits, 
l.h.s: action 1; r.h.s: action 2} \label{cir:3} 
\end{figure}

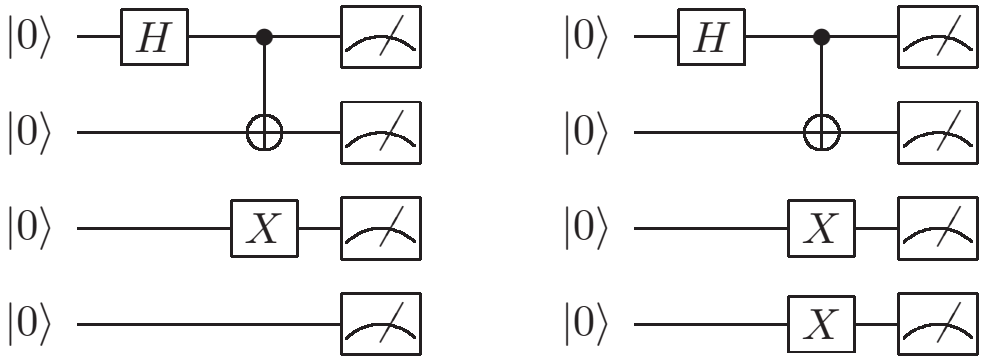
\begin{figure}[h!]
\centering
	\subfloat{
	\scalebox{1.5}{
		\begin{tabular}{c}
\Qcircuit @C=1em @R=.8em
{\lstick{\ket{0}}  & \gate{H} & \ctrl{1} & \meter \\
\lstick{\ket{0}}  & \qw & \targ & \meter  \\
\lstick{\ket{0}}  & \qw & \gate{X} & \meter  \\
\lstick{\ket{0}}  & \qw & \qw & \meter}  
		\end{tabular}
}
}
	\qquad \qquad
	\subfloat{
		\scalebox{1.5}{
		\begin{tabular}{c}
\Qcircuit @C=1em @R=.8em
{\lstick{\ket{0}}  & \gate{H} & \ctrl{1} & \meter \\
\lstick{\ket{0}}  & \qw & \targ & \meter  \\
\lstick{\ket{0}}  & \qw & \gate{X} & \meter  \\
\lstick{\ket{0}} &\qw  & \gate{X} &  \meter} 
		\end{tabular}
}
}
\caption{Corresponding quantum circuits, 
l.h.s: action 3; r.h.s: action 4} \label{cir:4} 
\end{figure}

We continue by defining the four obvious actions and their implementation. Action 1 
is merely not taking anything. The resulting state therefore is
\begin{equation}
|\psi_1\rangle = U_W (\ket{00} \otimes \ket{00})=
U_W^{[01]} \ket{00} \otimes \ket{00}.
\end{equation}
Carrying a jacket but no umbrella defines action 2. Hence the state is transformed as
\begin{equation}
|\psi_2\rangle = U_W \, \sigma_x^{[3]} (\ket{00} \otimes \ket{00})=
U_W^{[01]} \ket{00} \otimes \sigma_x^{[3]} \ket{00}.
\end{equation}
Action 3 is defined as taking an umbrella and no jacket; the final state follows as
\begin{equation}
|\psi_3\rangle = U_W \, \sigma_x^{[2]} (\ket{00} \otimes \ket{00})=
U_W^{[01]} \ket{00} \otimes \sigma_x^{[2]} \ket{00}.
\end{equation}
Of course one can take an umbrella as well as a jacket - defining action 4.
Obviously, it corresponds to the final state
\begin{equation}
|\psi_4\rangle = U_W \, \sigma_x^{[2]} \, \sigma_x^{[3]} (\ket{00} \otimes \ket{00})=
U_W^{[01]} \ket{00} \otimes (\sigma_x^{[2]} \otimes  \sigma_x^{[3]})\ket{00}.
\end{equation}
It is clear that for actions 1-4 the weather two-qubit state is entangled but that the remaining two qubits
are separable. The quantum circuits are shown in Figures (\ref{cir:3}, \ref{cir:4}).

\begin{figure}[h!]
\centering
	\scalebox{1.7}{
		\begin{tabular}{c}
\Qcircuit @C=1.em @R=.8em
{\lstick{\ket{0}} & \gate{H} & \ctrl{1} & \qw      & \ctrl{2} & \gate{H} & \ctrl{1} & \qw & \meter \\
\lstick{\ket{0}}  & \qw      & \targ    & \ctrl{2} & \qw      & \qw & \targ &  \qw & \meter\\ 
\lstick{\ket{0}}  & \qw      & \qw      & \qw      & \targ    & \qw & \qw &  \qw & \meter\\
\lstick{\ket{0}}  & \qw      & \qw      & \targ    & \qw      & \qw & \qw &  \qw & \meter 
\gategroup{1}{4}{4}{5}{1em}{--}
}
		\end{tabular}
}
\caption{Quantum circuit for action 5, the CNOTs in the box can be done simultaneously}  \label{cir:5} 
\end{figure}
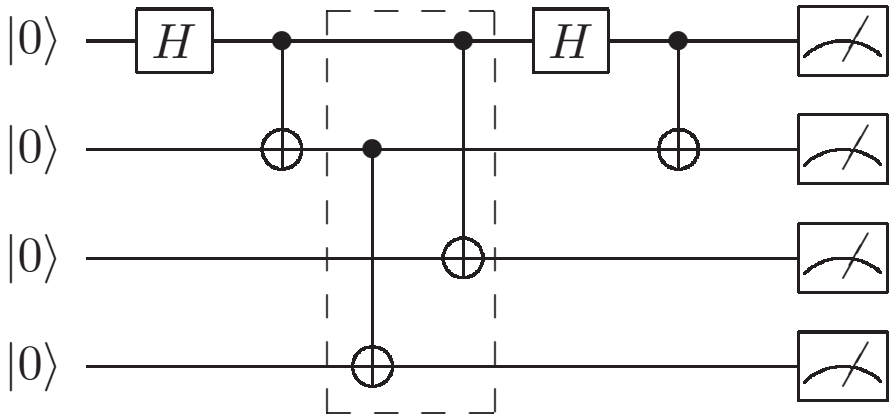

Once again, an alternative action 5 is obtained by
 modeling an intermediate decision moment.
It depends on the actual weather condition and is implemented by a CNOT operation. 
Explicitly, we then obtain  
\begin{equation}
|\psi_5\rangle = U_W \, C_N^{[13]}\,  C_N^{[02]}\,  U_W \,  (\ket{00} \otimes \ket{00}).
\end{equation}
The quantum circuit is depicted in Figure (\ref{cir:5}).

For action 5 there is an additonal cost $d$ leading to
\begin{equation}
c_5(s_0,s_1)=c_0 s_2 + c_1 s_3 +d.
\end{equation}
This action and its implementation may be called `with experimentation'.

\subsection{Results}
The quantum mechanical calculations in this extended example are still analytically feasible.
As above, we supplement and confirm these by the use of the QX simulator. For more complicated use 
cases, analtyical predictions are impossible and one has to rely on the QX simulator only.
Eventually, the aim is to get samples from the NISQ computer. 

We present the results for each action in terms of probabilities for obtaining the
classical bits $\vec s$ and the eventual expected value for the reward.
Note that we only list the non-zero probabilities
\begin{itemize}
\item{Action 1} 
\begin{equation}
p(0,0,0,0)= p(1,1,0,0)= \tfrac{1}{2}, \quad E_1[r]= \tfrac{1}{2}(v_0 + v_1).
\end{equation}
\item{Action 2} 
\begin{equation}
 p(0,0,0,1)= p(1,1,0,1)= \tfrac{1}{2}, \quad E_2[r]= \tfrac{1}{2}v_0 +v_1-c_1.
\end{equation}
\item{Action 3}
\begin{equation}
p(0,0,1,0)= p(1,1,1,0)= \tfrac{1}{2}, \quad E_3[r]= \tfrac{1}{2}v_1 +v_0-c_0.
\end{equation}
\item{Action 4}
\begin{equation}
p(0,0,1,1)= p(1,1,1,1)= \tfrac{1}{2}, \quad E_4[r]= v_1 +v_0-c_0-c_1.
\end{equation}
\item{Action 5} 
\begin{eqnarray}
p(0,0,0,0)&=& p(1,1,0,0)= p(0,1,1,1)= p(1,0,1,1)= \tfrac{1}{4}, \nonumber \\
E_5[r] &=& \tfrac{3}{4}(v_0 +v_1)-\tfrac{1}{2}(c_0+c_1)-d.
\end{eqnarray}
\end{itemize}
In principle, there are no free parameters and one just has to compare the expected 
rewards in order to take the optimal decision. It can be shown that for $d > 0$, action 5 is never
the best action. 

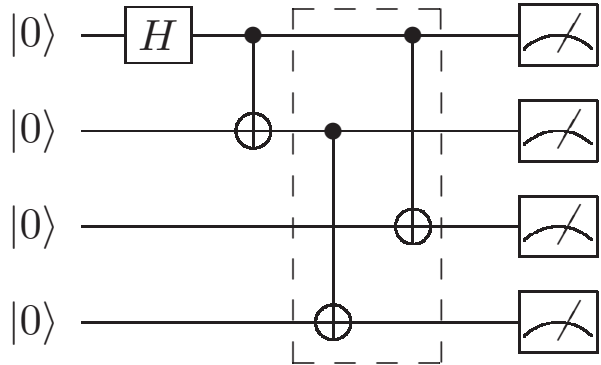
\begin{figure}[h!]
\centering
	\scalebox{1.5}{
		\begin{tabular}{c}
\Qcircuit @C=1.em @R=.8em
{\lstick{\ket{0}} & \gate{H} & \ctrl{1} & \qw      & \ctrl{2} &  \qw & \meter \\
\lstick{\ket{0}}  & \qw      & \targ    & \ctrl{2} & \qw      &  \qw & \meter\\ 
\lstick{\ket{0}}  & \qw      & \qw      & \qw      & \targ    &  \qw & \meter\\
\lstick{\ket{0}}  & \qw      & \qw      & \targ    & \qw      &  \qw & \meter 
\gategroup{1}{4}{4}{5}{1em}{--}
}
		\end{tabular}
}
\caption{Quantum circuit for action 6, the CNOTs in the box can be done simultaneously}  \label{cir:6} 
\end{figure}

Suppose it would be possible to postpone the intermediate decision to the moment of the 
truth\footnote{In the quantum model corresponding to the ideal measurement, that is 
instantaneous projection.}.
It is obvious that such an action profile gives better rewards than action 5 and plausible
that it is the best of all for small enough $d$. In order to check these staements,
we define the action 6
as action 5 without the last $U_W$ operation.
Its quantum circuit is shown in Figure (\ref{cir:6}).

\begin{figure}[htb]
\centering
\includegraphics[width=7.75cm]{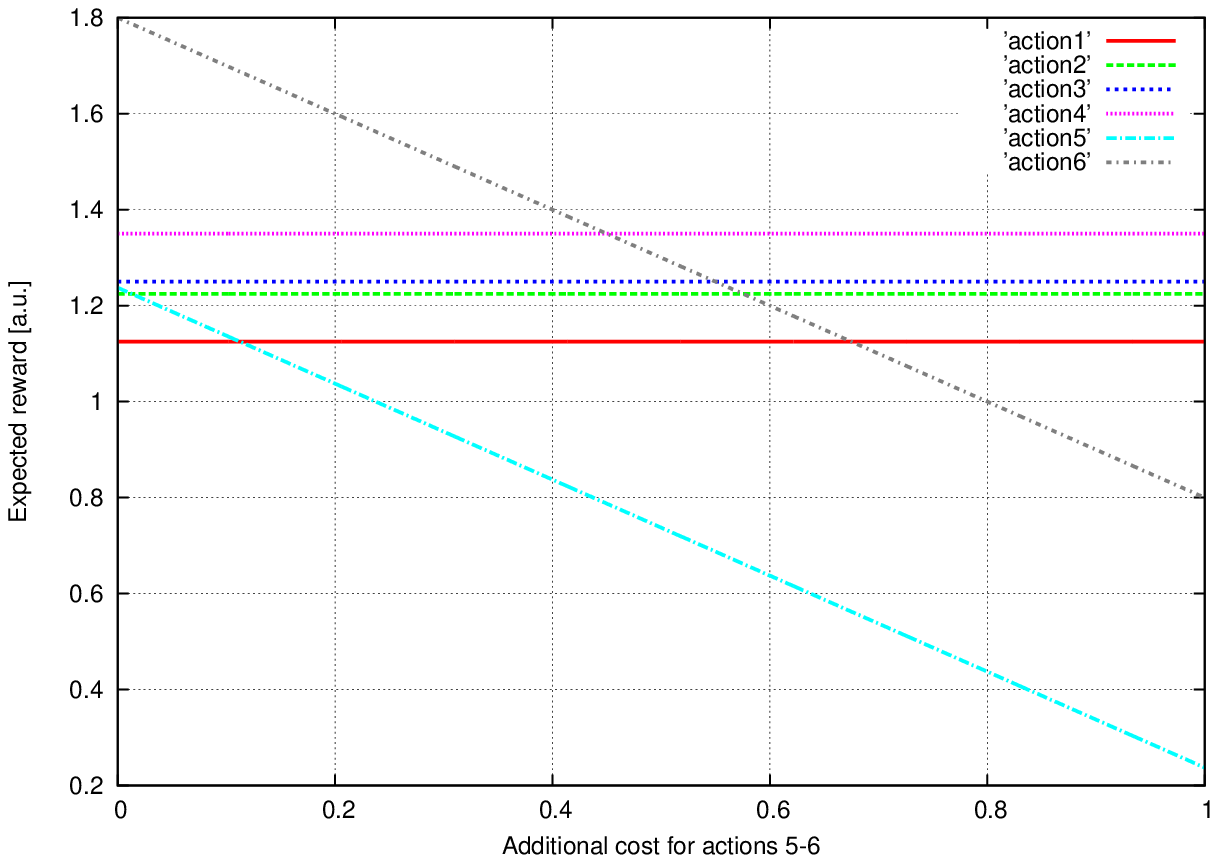}
\includegraphics[width=7.75cm]{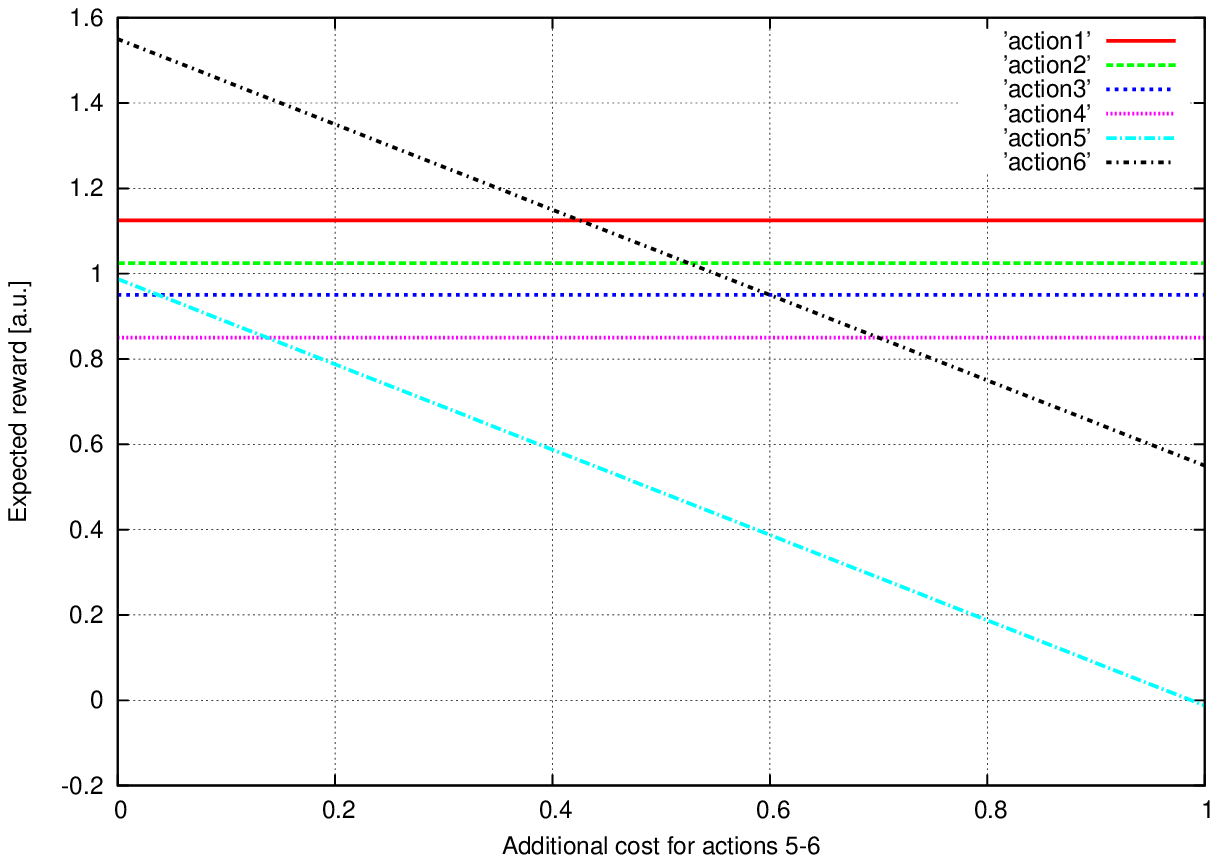}
\caption{Expected rewards for the six different actions as a function of the additional cost $d$.
L.h.s. $c_0=0.5, c_1=0.4 $; r.h.s $c_0=0.8, c_1=0.6 $.  \label{fig:3}}
\end{figure}

Thus we get as state
\begin{equation}
|\psi_6\rangle =  C_N^{[13]}\,  C_N^{[02]}\,  U_W \,  (\ket{00} \otimes \ket{00}).
\end{equation}
Of course, the additional cost $d$ has to be included in the reward.
The resulting probabilities and expected reward are obtained as
\begin{equation}
p(0,0,0,0)= p(1,1,1,1)= \tfrac{1}{2}, \quad E_6[r]= v_1 +v_0-\tfrac{1}{2}(c_0+c_1)-d.
\end{equation}
We illustrate typical outcomes in this example in Figure (\ref{fig:3}). The rewards for the
various actions are plotted as a function of the additional cost $d$ in actions 5-6. 
The value parameters are fixed as $v_0=1.25, v_1=1.0$.
 
\subsection{Including variances}
\begin{figure}[htb]
\centering
\includegraphics[width=7.75cm]{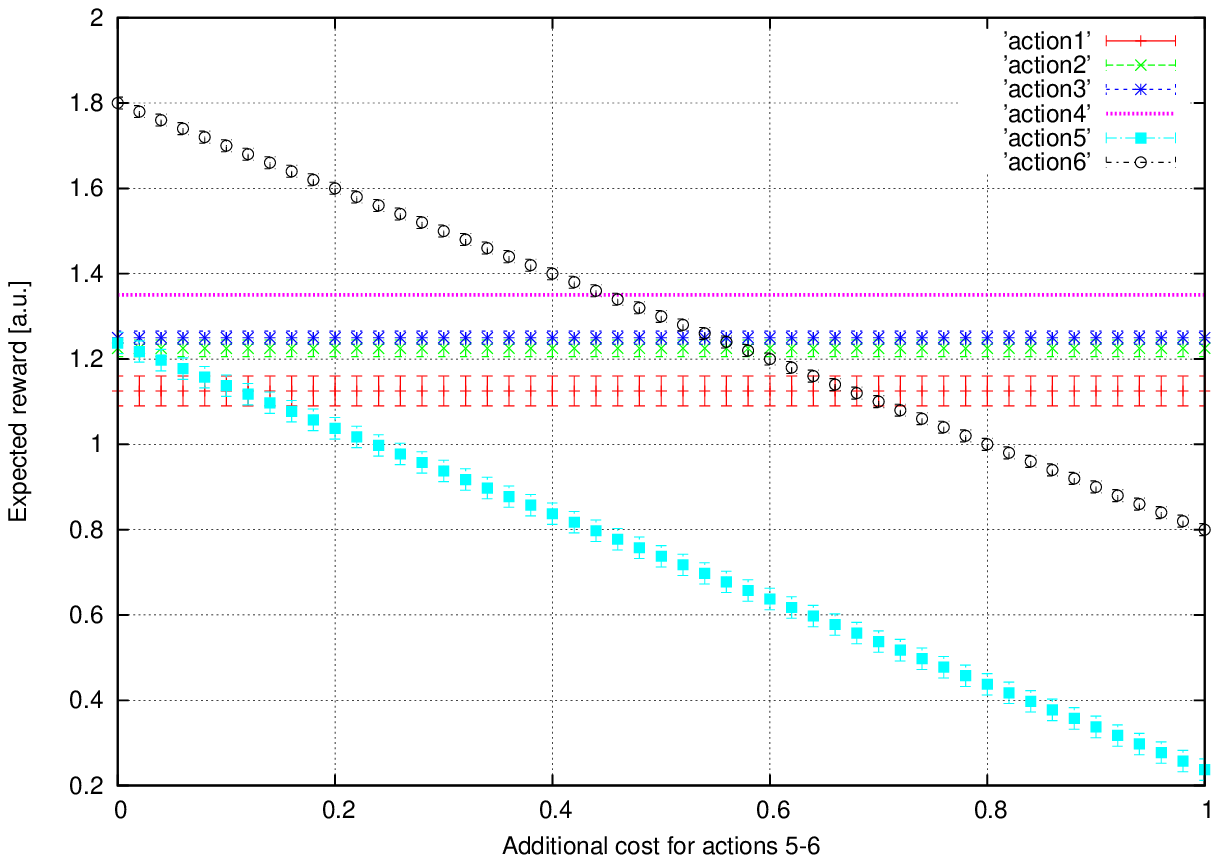}
\includegraphics[width=7.75cm]{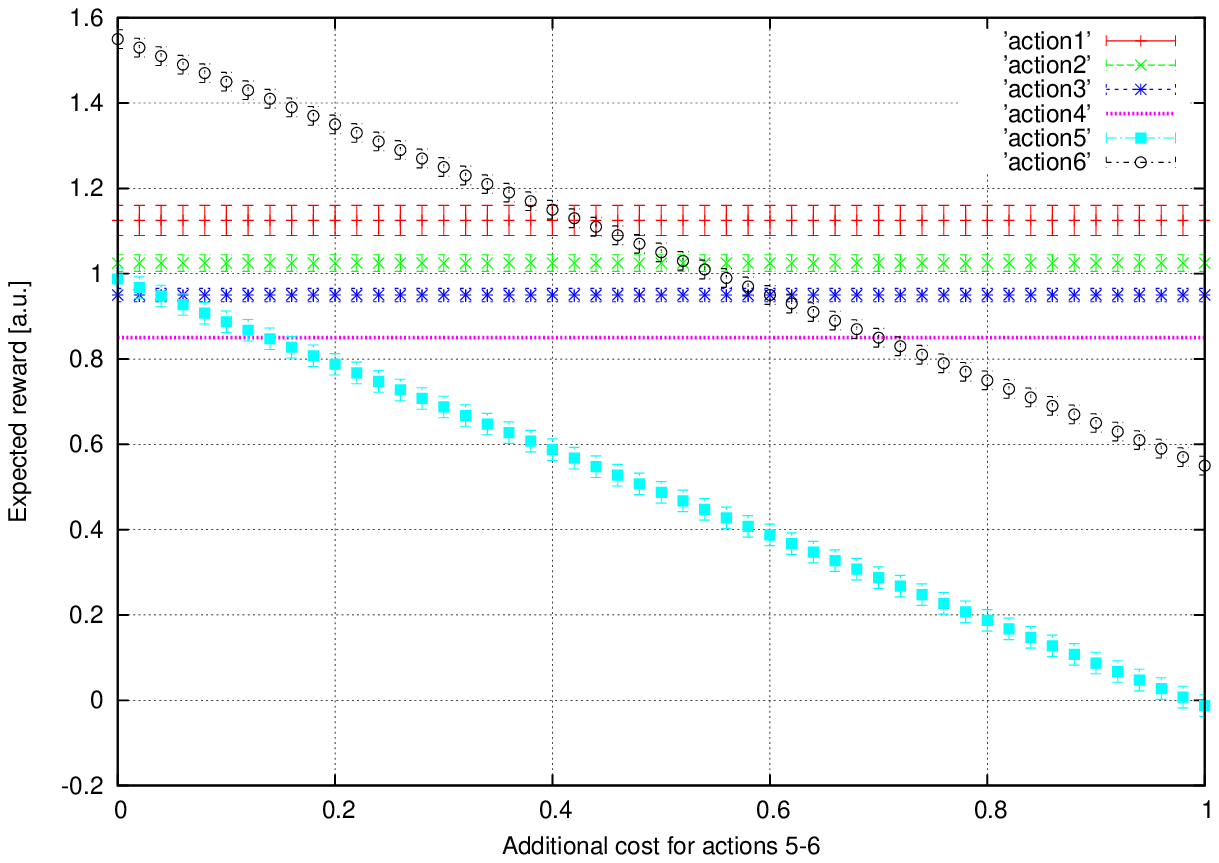}
\caption{Expected rewards and variances for the six different actions as a function of the additional cost $d$.
L.h.s. $c_0=0.5, c_1=0.4 $; r.h.s $c_0=0.8, c_1=0.6; N=1024 $.  \label{fig:4}}
\end{figure}

The variances and standard deviations of the rewards
can also be straightforwardly calculated for the various actions. 
The following results are obtained
\begin{eqnarray}
\sigma_1 &=& \tfrac{1}{2}(v_0+v_1), \quad \sigma_2 =\tfrac{1}{2}v_0, \quad
\sigma_3 = \tfrac{1}{2}v_1, \quad \sigma_4 =0, \nonumber \\
\sigma_5 &=& \tfrac{1}{2} \sqrt{\tfrac{3}{4}(v_0+v_1)^2+(c_0+c_1)^2-(v_0+v_1)(c_0+c_1)}, \nonumber \\
\sigma_6 &=& \tfrac{1}{2}(c_0+c_1).
\end{eqnarray}
Also in this example the standard deviations are of the same order of magnitude as the rewards.
It prohibits decision making based on one run. As above, we assume that $N=1024$ runs are 
performed which reduces the standard deviation by $1/\sqrt{N}$. 
The expected rewards including these standard deviations are depicted in Figure (\ref{fig:4})
	
\section{Including noise in the QX simulator}
At present, only one error model is available in the QX simulator in Quantum Inspire,
that is the ``symmetric depolarizing channel" \cite{QI}. The per-operation
error probability has to be set; typical values are between 0.001 and 0.01. 
In order to get a first assessment of the consequences, we have performed some
additional simulations for action 6 of the second example. 
The parameters are chosen as in the l.h.s. of Figure (\ref{fig:3}); in addition we fix the
additional cost as d=0.5. The number of runs remains $N=1024$. The theoretical result for
the expected reward and its standard deviation is
$E_6[r] = 1.30 \pm 0.014$. 
We compare the sample mean and corresponding estimated error to theory
for three error probabilities.
In the noiseless case, we obain the same result, for 
error probability 0.001 we get $1.33 \pm 0.014$, whereas
in case of error probability 0.01 we find $1.29 \pm 0.024$.
Of course, also these values slightly vary repeating the simulations of 1024 runs.
These first results, only based on the implemented error model in the
QX simulator, indicate that noise does not necessarily
prohibit decision making using a NISQ computer. 
There will be a noise limit/threshold of course. Here we have seen that 
for an error probability  of 0.1, results appear completely random.

\section{Conclusion and outlook}
An approach for decision making under uncertainty using a NISQ computer is proposed. 
According to Bayes decision rule \cite{OR}, the decision alternative with the maximum expected reward
is chosen. The reward is defined as the difference between value and cost and these
are supposed to be formulated as functions of classical binary variables. 
Each binary variable requires a qubit in a NISQ device. Measuring these qubits in the standard
basis generates a sample of the binary variables. Because of the inherent probabilistic
nature of such a quantum measurement, the samples are random thereby reflecting the
uncertainty in the decision problem. The noise in a NISQ computer will also add to this. 
The various decision actions and the development of the `state of nature' are modelled as unitary
transformations on the qubits. After initialization, these operations are to be
performed on the qubits before measuring them. Such a quantum program has to run a sufficient
number of times. 

The idea is demonstrated by means of some simple examples of decision making.
In fact, the calculations can still be performed analytically. Nevertheless we have
cross-checked results by generating qubit measurement outcomes to obtain samples.
Awaiting a NISQ device, we have actually used the QX simulator of Quantum Inspire\cite{QI}.
The results indicate the feasibility of the approach. In order to reduce the variance
in the expected rewards, multiple runs of the quantum algorithm are necessary.
The QX simulator supports 1024 runs, which is sufficient for the problems considered.

In classical decision making in the face of uncertainty, one relies on probabilities
and probability density functions. Judicious choices for such distributions and classical
sampling and/or probability theory computations are necessary. In the proposed quantum approach
this is replaced by unitary transformations and measurement of the qubits. Note that 
the formulation of the classical reward function remains in principle the same.
It is constrained, however, by its dependence on random binary variables.

It is planned to test these ideas as soon as a up-and-running NISQ computer is available.
A small device is already scheduled for this year \cite{QI}.
If these first evaluations are succesful, then larger and more complex decision making
problems may be implemented. 
\section*{Acknowledgements}
The author thanks Kelvin Loh for a critical reading of the manuscript and 
J\'er\'emy Veltin for his support.

\end{document}